\documentclass[aps,prl,superscriptaddress,longbibliography,twocolumn
] {revtex4-2}
\usepackage{graphicx}
\usepackage{epstopdf}
\usepackage{bm}
\usepackage{amsmath}
\usepackage{verbatim}
\usepackage{amssymb}
\usepackage{listings}
\usepackage{color}
\usepackage{cases}
\definecolor{darkblue}{rgb}{0,0,0.6}
\usepackage[colorlinks,linkcolor=darkblue,citecolor=darkblue,urlcolor=darkblue]
{hyperref}
\usepackage{xspace}
\usepackage{ulem}
\usepackage{wasysym}
\usepackage{mathtools}
\usepackage{stmaryrd}
\usepackage{marvosym}
\usepackage{stackengine}
\usepackage{setspace}
\usepackage{epsfig,verbatim}

\newcommand\subfig[2]{{Fig.~\ref{#1}{#2}}}
\newcommand\subcap[1]{{(#1):}}
\newcommand{\REF}[2][]{
        \ifthenelse{\equal {#1} {}}{Ref.~\cite{#2}}{Ref.~\cite[#1]{#2}}}
\renewcommand{\emph}[1]{\textit{#1}}

\newcommand{\vpointer}{v_{\vv{}}}

\newcommand{\alphacrit}{\alpha_\text{crit}}
\newcommand{\liftedconf}{c}

\usepackage{ifthen}
\usepackage{float}
\usepackage{esvect}
\newcommand{\eq}[1]{eq.~(\ref{#1})}

\newcommand{\eqfromto}[2]{eqs.~(\ref{#1}) to~(\ref{#2})}

\newcommand{\fig}[1]{Fig.~\ref{#1}}
\newcommand{\quot}[1]{``#1''}

\newcommand{\etc}{\textrm{etc.}}

\newcommand{\ie}{\textrm{i.e.}}

\newcommand{\CCAL}{\mathcal{C}}  
\newcommand{\rhobar}{\overline{\rho}}  

\newcommand{\mean}[1]{\left\langle #1 \right\rangle}
\newcommand{\half}{\frac{1}{2}}

\newcommand{\TASEP}{\protect \textsc{Tasep}\xspace}
\newcommand{\LT}{\protect lifted \textsc{Tasep}\xspace}
\newcommand{\LUT}{\protect lifted-\textsc{Tasep}\xspace}
\newcommand{\TSAW}{TSAW\xspace}

\newcommand{\HOLE}{\fbox{$\circ$}}
\newcommand{\LEFT}{\fbox{$\mathrlap{\mkern-5mu \leftarrow}\circ$}}
\newcommand{\RIGHT}{\fbox{$\mathrlap{\mkern-5mu \rightarrow}\bullet$}}
\newcommand{\PART}{\fbox{$\bullet$}}

\newcommand{\ZERO}{\fbox{\phantom{$\bullet$}}}
\newcommand{\ONE}{\fbox{$\bullet$}}

\newcommand{\TWO}{\fbox{$\mathrlap{\mkern-4mu \rightarrow}\bullet$}}

\def\dosquare{\kern-.1em\tikz[overlay]\draw[red,rounded
corners,fill=red!15!white,baseline] (0,1.9ex) rectangle
(2.4em,-3.2ex);\kern.1em}

\newcommand{\TSAWfig}[1]{\TSAW\nolinebreak\!\!(#1)\xspace}

\newcommand{\rhohigh}{\rho^{\text{high}}}
\newcommand{\rholow}{\rho^{\text{low}}}
\newcommand{\rhohighlow}{\rho^{\text{high,low}}}
\newcommand{\rholocal}{\rho_\ell}
\newcommand{\critsize}[1]{|\CCAL_{#1}|}
\newcommand{\crit}[1]{\CCAL_{#1}}
\newcommand{\VLT}{nearest-neighbor lifted \textsc{Tasep}\xspace}
\newcommand{\teq}{t_{\text{eq}}}

\begin{document}

\title{
Velocity trapping in the lifted TASEP  and the true self-avoiding random walk
}

\author{Brune Massouli\'e}
\affiliation{CEREMADE, CNRS, Universit\'e Paris-Dauphine, Universit\'e PSL,
75016 Paris, France}

\author{Cl\'ement Erignoux}
\affiliation{Inria MUSICS, ICJ UMR5208, CNRS, Ecole Centrale de Lyon, INSA Lyon,
Universit\'e Claude Bernard Lyon 1, Universit\'e Jean Monnet, 69603
Villeurbanne,
France}
\author{Cristina Toninelli}
\affiliation{DMA, ENS Université PSL, 45 rue d'Ulm 75005 Paris, France}
\affiliation{CEREMADE, CNRS, Universit\'e Paris-Dauphine, Universit\'e PSL,
75016 Paris, France}
\author{Werner Krauth} \affiliation{Laboratoire de Physique de l’Ecole normale
sup\'erieure, ENS,
Universit\'e PSL, CNRS, Sorbonne Universit\'e, Universit\'e de Paris Cit\'e,
Paris, France}
\affiliation{Rudolf Peierls Centre for Theoretical Physics, Clarendon
Laboratory, Oxford OX1 3PU, UK}
\affiliation{Simons Center for Computational Physical Chemistry, \\
New York University, New York (NY), USA}

\date{\today} 

\begin{abstract}
We discuss non-reversible Markov-chain Monte Carlo algorithms that, for
particle systems, rigorously sample the positional Boltzmann distribution
and that have faster than physical dynamics. These algorithms all feature a
non-thermal velocity distribution. They are exemplified by
the \LT (totally asymmetric simple exclusion process), a one-dimensional lattice
reduction of event-chain Monte Carlo. We analyze its dynamics
in terms of a velocity trapping that arises from correlations between the local
density and the particle velocities. This allows us to formulate a conjecture
for its out-of-equilibrium mixing time scale, and to rationalize its equilibrium
superdiffusive time scale. Both scales are faster than for the (unlifted)
\TASEP.
They are
further justified by our analysis of the \LT in terms of many-particle
realizations of true self-avoiding random walks. We discuss velocity trapping
beyond the case of one-dimensional lattice models and in more than one physical
dimensions. Possible applications beyond physics are pointed out.
\end{abstract}

\maketitle
All of statistical mechanics descends from the insight that the  equilibrium
motion of atoms in a classical gas,  liquid, or solid, \etc, obeys the
distribution obtained by Maxwell in 1859 ~\cite{Brush1976}. For two phases of a
material (say, a gas and a crystal) in thermal contact in a sample, atoms thus
have the same instantaneous velocity distribution independent of (decoupled
from) whether they weakly interact in the gas, or are tightly confined in the
crystal lattice. The decoupling  of the kinetic from the potential degrees of
freedom underlies the microscopic interpretation of the temperature.

In computational physics, the decoupling of velocities and positions is
ubiquitous. In the vast field of classical molecular
dynamics~\cite{SchlickBook,FrenkelSmitBook2001,Ciccotti2022}, for example, the
particle velocities in a finite system evolve with the gradient of the
potential, but are furthermore updated from a thermostat, \ie, a Gaussian
distribution specified by the system temperature. In Hamiltonian Monte
Carlo~\cite{Duane1987,Neal2011}, a rigorous version of molecular dynamics, all
particle velocities are periodically resampled from the Maxwell distribution in
a way that is blind to its environment. The independence of positions and
velocities is echoed in Monte Carlo methods~\cite{SMAC} that at each time step
attempt to move a random particle.

In recent years, non-reversible event-chain Monte Carlo methods
\cite{Bernard2009,Michel2014JCP,Krauth2021eventchain} based on lifted Markov
chains \cite{Diaconis2000,Chen1999} have featured faster convergence than
possible under physical dynamics~\cite{Bernard2011,Kampmann2021,Klement2019}.
They exactly sample the Boltzmann distribution of the positions. Velocities,
generalized into lifting variables, do not follow the Maxwell distribution, and
they correlate with local particle densities. In the present paper, we analyze
the superdiffusive convergence of these algorithms. We justify the observed
link~\cite{Maggs2024,Maggs2024b} between one-dimensional event-chain
algorithms~\cite{KapferKrauth2017,Lei2019} and true self-avoiding random walks
(\TSAW)~\cite{Amit1983}, that were much studied in
physics~\cite{Pietronero1983,BernasconiPietronero1984,Bremont2024} and in
mathematics~\cite{Toth1995,TothWerner1998,Veto2008,Dumaz2013}.
\begin{figure*}[t!h]
    \centering
    \includegraphics[width=2.0 \columnwidth]{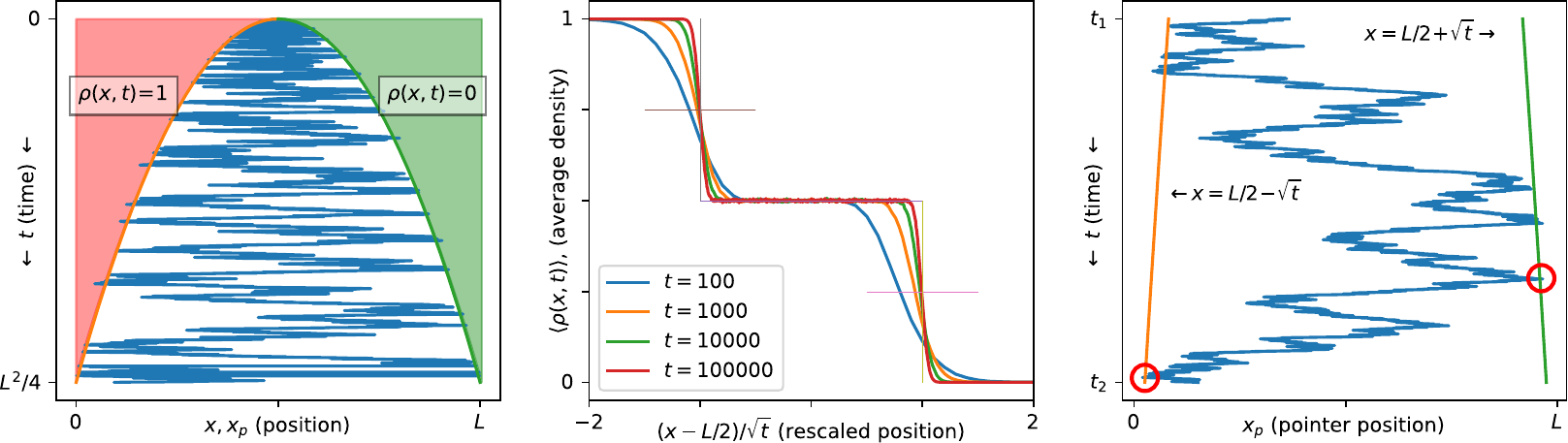}
    \caption{Time evolution of the \LT  ($N=10^6, L=2N)$ at $ \alpha  =
\alphacrit = \half$, from an initial \quot{step} (at $t=0$, $\rho(x<
L/2)=1$ and $\rho(x> L/2)=0$). \subcap{a} Pointer position $x_{\text{p}}$
\emph{vs} time $t$, showing the dome-shaped critical region $\crit{t}$
expanding at the expense of the step. \subcap{b} Ensemble-averaged density
$\mean{\rho(x,t)}$, for rescaled positions $(x - L/2)/\sqrt{t}$ (the pointer
position is excluded from the histogram of densities). The critical region
appears clearly, and the ensemble-averaged interfaces
have width $\sim t^{1/4}$. \subcap{c} Close-up of trajectory, between boundaries
$L/2 \pm \sqrt{t}$ of the critical region. A pair of events defining a transfer
is highlighted.}
\label{fig:Triptych}
\end{figure*}

For concreteness, we consider the \LT~\cite{Essler2024PRX} (totally
asymmetric simple exclusion process), a non-reversible Markov chain for $N$
hard-sphere particles on a periodic one-dimensional $L$-site lattice.
In the \LT, a single particle carries a pointer and can move, but the
dynamics also determines which particle can move next:
\begin{align}
\hspace{-1.3cm}
\underbrace{
\overset{k}{\ONE} \overset{l}{\TWO} \ZERO \overset{m}{\ONE}}_{\liftedconf_t}
&\rightarrow \hspace{-0.1cm}
 \underbrace{\overset{k}{\ONE} \ZERO \overset{l}{\TWO}
\overset{m}{\ONE}}_{\text{\quot{move} (determ.)}} &
\hspace{-1.5cm}
\rightarrow 
\underbrace{
\begin{cases}
 \overset{k}{\TWO} \ZERO \overset{l}{\ONE} \overset{m}{\ONE} & \quad  \alpha
\hspace{-1cm}\\
 \ONE \ZERO \TWO  \ONE & 1-\alpha \hspace{-1cm}
\end{cases}}_{\liftedconf_{t+1}}
\label{equ:LTASEP1}
 \\
\hspace{-1.3cm}
\underbrace{
\overset{k}{\ONE} \overset{l}{\TWO} \overset{m}{\ONE} \ZERO}_{\liftedconf_t}
&\rightarrow \hspace{-0.1cm}
\underbrace{\overset{k}{\ONE} \overset{l}{\ONE} \overset{m}{\TWO}
\ZERO}_{\!\!\!\!\text{\quot{collide}
(determ.)}\!\!\!\!}  &
\hspace{-1.5cm}
\rightarrow 
\underbrace{
\begin{cases}
 \overset{k}{\ONE} \overset{l}{\TWO} \overset{m}{\ONE} \ZERO  &\quad \alpha
\hspace{-1cm} \\
 \ONE \ONE \TWO \ZERO  &1-\alpha \hspace{-1cm}
\end{cases}
}_{\liftedconf_{t+1}}
\label{equ:LTASEP2}
\end{align}
Between times $t$ and $t+1$, the active particle (here $l$)  first moves to its
right, keeping the pointer (\eq{equ:LTASEP1}) or, if this is not possible, first
collides with its right-hand neighbor $m$, passing the pointer on to it
(\eq{equ:LTASEP2}). Then, in both cases, with probability $\alpha>0$, the
pointer is pulled back from the active particle to its left-hand neighbor (from
$l$ to $k$ or from $m$ to $l$). With periodic boundary conditions, the above
dynamics converges towards a stationary state with, for any pullback $\alpha$,
equal weights for all lifted configurations $\liftedconf$ (positions of
particles and of the pointer)~\cite{Essler2024PRX}.

In the \LT, only the active particle is responsible for mass transport which, in
one time step, equals either one (as in \eq{equ:LTASEP1}) or zero (as in
\eq{equ:LTASEP2}). The pullback $\alpha$ decouples the mass transport from the
pointer drift $\vpointer$, the difference between pointer positions in
$\liftedconf_{t+1}$ and $\liftedconf_t$ (periodic boundary conditions being
accounted for). Two averages of the pointer drift $\vpointer$ are relevant:
\begin{equation}
\underbrace{
\mean{\vpointer} =
1-\frac{\alpha}{\rhobar}}_{\!\!\!\!\!\text{equilibrium-averaged, $\rhobar =
N/L$}\!\!\!\!\!\!}
\; \mbox{ and }\; 
 \underbrace{\vpointer(\rholocal, \alpha) =
\frac{\rholocal-\alpha}{\rholocal}}_{\text{coarse-grained}}.
\label{equ:PointerVelocity}
\end{equation}
On the left-hand side of \eq{equ:PointerVelocity}, the equilibrium-averaged
pointer drift is expressed in terms of the system density $\rhobar = N/L$, and
it rigorously vanishes for $\alpha= \alphacrit = \rhobar$~\cite{Essler2024PRX}.
The expression on the right-hand-side of \eq{equ:PointerVelocity} translates
this relation to the case of a pointer inside a local region of suitably chosen
size $\ell$ with a coarse-grained local density $\rholocal$. For the critical
pullback $\alphacrit$, the pointer drift
\begin{equation}
\vpointer(\rholocal, \alphacrit) =
\frac{\rholocal-\alphacrit}{\rholocal} =
\frac{\Delta \rholocal}{\rholocal}
\label{equ:PointerVelocityCrit}
\end{equation}
exposes a linear coupling to the local excess density $\Delta \rholocal =
\rholocal - \rhobar$.

The pointer-drift--density coupling of \eq{equ:PointerVelocityCrit} contains the
mechanism that traps the pointer. This is most notable when starting the \LT
at $\alphacrit$
from a \quot{step} initial configuration at time $t=0$ with the left-hand side
of the system at high local density $\rhohigh > \rhobar$, and the right-hand
side at low density $\rholow < \rhobar$ (see \subfig{fig:Triptych}{a}). From an
initial position $x$ in the high-density region (where $\vpointer>0$) the
pointer is expelled towards the right, and from a low-density region it is
expelled towards the left. Trapped in the middle, the pointer moves back and
forth, creating a dome-shaped critical region $\crit{t}$,
which expands while maintained in
equilibrium at density $\sim \rhobar$. The scaling behavior of the
ensemble-averaged density $\rho(x,t)$ starting from the
step initial configuration defines $\crit{t}$ precisely
as the region with an asymptotic
density $\rhobar$ (see \subfig{fig:Triptych}{b}). Although the ensemble average 
represented in \subfig{fig:Triptych}{b} spreads the position of the interface by 
$\sim t^{1/4}$ around its expected position, the critical region is
bounded by sharp interfaces, which trap the pointer through the mechanism of
\eq{equ:PointerVelocity}. Its size $\critsize{t}$ for the
step initial configuration with $\rhohighlow = (1,0)$ (the case shown in
\fig{fig:Triptych}) follows from assuming the pointer to be
uniformly distributed in $\crit{t}$. Since a unit mass transport
(\eq{equ:LTASEP1}) occurs with probability $\frac12$, each point of the critical
region moves forward by $\frac{1}{2\critsize{t}}$ in one step, expanding it
in both directions (because the leftmost particle pushes an empty
site backwards):
\begin{equation}
\critsize{t+1}
 \simeq \critsize{t} +  \frac{1}{\critsize{t}}.
 \label{equ:Dome10bis}
\end{equation}
This difference equation for $\critsize{t}$ (with $\critsize{0} \sim 1$) is
solved by $\critsize{t} = 2 \sqrt{t}$ for large $t$, an expression
well confirmed for the single trajectory of \subfig{fig:Triptych}{a} and
for the ensemble average of trajectories of \subfig{fig:Triptych}{b}.

\begin{figure*}[t!h]
    \centering
    \includegraphics[width=2.0 \columnwidth]{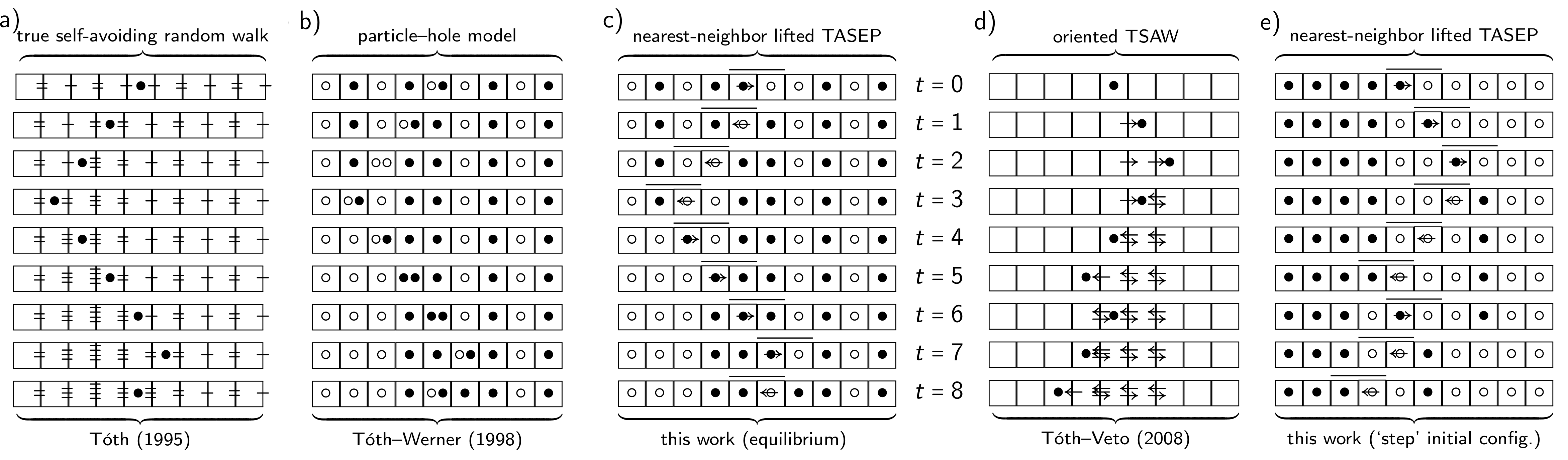}
    \caption{Correspondence of \TSAW's with particle models related  to the \LT.
\subcap{a} Zero-temperature version of the \TSAW introduced in \REF{Toth1995}, with local
times indicated by horizontal lines, initialized as in \REF{TothWerner1998}.
\subcap{b} Particle--hole representation of (a)~\cite{TothWerner1998}. From the
doubly occupied site, a particle  moves to the right, or a hole to the left.
\subcap{c} Equivalent representation of (a) in terms of the \VLT of
\eqfromto{equ:Fleches1}{equ:Fleches3}. \subcap{d} Oriented \TSAW with
directional local times as initialized in \REF{Veto2008}. \subcap{e} Equivalent
representaton of (d) in terms of the \VLT, with a \quot{step} initial
configuration.}
\label{fig:TSAW}
\end{figure*}

The  analysis of \eq{equ:Dome10bis} generalizes to a
density step $\rhohighlow = \rhobar \pm \varepsilon$, resulting in
\begin{equation}
\critsize{t} = 2 \sqrt{t/(2\varepsilon)}\quad (\rhohighlow= \rhobar \pm
\varepsilon).
\label{equ:ExpandingDomeEpsilon}
\end{equation}
This dimensionally corrected version of the solution to \eq{equ:Dome10bis} is
backed up by numerical solutions analogous to
\subfig{fig:Triptych}{b}~\cite{Erignoux2025b}. It incorporates that any excess
particle, at density $\rhobar$, must be spread out over a distance $\sim
1/\varepsilon$ if the initial density beyond the right boundary is
$\rhobar-\varepsilon$. In equilibrium, step-like initial configurations with
$\varepsilon \sim~ L^{-1/2}$ arise from thermal particle-number fluctuations.
From \eq{equ:ExpandingDomeEpsilon}, the time scale on which such a step spreads
to the system-size scale $L$ is given by
\begin{equation}
\teq \sim  L^{3/2}.
\label{equ:RelaxationTime}
\end{equation}
The equilibrium fluctuations in the $L$-site \LT are thus expected to reset on a
superdiffusive time scale  $L^{3/2}$ that naturally arises from the length scale
$L$ together with the velocity scale $L^{-1/2}$ of \eq{equ:PointerVelocityCrit}
for equilibrium density fluctuations. The  time scale $\teq$ is a prime
candidate for the relaxation time (the inverse absolute gap) of the \LT, and it
is readily seen in numerical
simulations~\cite{Lei2019,Essler2024PRX,Krauth2024hamiltonian}. However,
numerical Bethe-ansatz solutions of the \LT transition
matrix~\cite{Essler2024PRX,Essler2025} point to the existence of a (possibly
isolated) eigenvalue with a spectral gap scaling as $1/L^2$ and suggesting that,
strictly speaking,  the relaxation time of the \LT (for $N \propto L$ and
$\alpha = \alphacrit$) may be $\sim L^2$. Up to this possible subtlety, we
conjecture that in equilibrium not only the pointer travels through the system
$L$ in time $\sim L^{3/2}$ (and likewise, through the critical region in time
$\sim \critsize{t}^{3/2}$), but also that each of its passes through the lattice
produces an essentially independent equilibrium configuration.

The microscopic mechanism for the expansion of the critical region relies on a
subtle interplay of equilibrium particle-number fluctuations with the pointer
drift. For a single trajectory starting from the step initial configuration we
treat the case $\alpha = 1/2$, $\rhohighlow = (1,0)$ (see
\subfig{fig:Triptych}{c}). At a given time $t \lesssim L^2/4$, the size of the
critical region is $\critsize{t} \sim \sqrt{t}$. Our simulations confirm that
the pointer requires $ \delta t\sim \critsize{t}^{3/2}$ steps to move across the
critical region, which is consistent with it being maintained in equilibrium,
driving the pointer in either direction through the effect of thermal
fluctuations. During each crossing, mass transport advances the particles in
$\crit{t}$ by $\delta n\sim \critsize{t}^{3/2}/\critsize{t} = \critsize{t}^{1/2}
$, expanding the critical region by the same order. A positive density
fluctuation, for example, drives the pointer away from the left boundary and
towards the right. It induces mass transport, which ultimately spreads the
critical region towards the low-density region by $\sim \critsize{t}^{1/2}$,
thus lowering the density  and eventually reversing the sign of
the density fluctuation in $\crit{t}$. This then drives the pointer back to the left,
towards the high-density region. Through each visit of the pointer at  the left
boundary of $\crit{t}$,  mass is pushed forward, spreading $\crit{t}$ to the
left. Eventually, $\crit{t}$ grows to the left by  $\sim
\critsize{t}^{1/2}$, which once again reverses the density fluctuations. In
order for $\crit{t}$ to grow macroscopically (of order $\critsize{t}$), 
this back-and-forth motion takes
place $\sim \critsize{t}^{1/2}$ times. Each pointer crossing of the critical
region requires a time $\sim \critsize{t}^{3/2}$ and it ultimately reverses the
fluctuation, thus resulting in a macroscopic growth time scale $\critsize{t}^2$.
This mechanism is consistent with our previous estimate $\delta t\sim
\critsize{t}^{3/2}$ and $\delta n\sim \critsize{t}^{1/2} $, and it supports the
picture of the critical region in effective equilibrium.

The \LT is closely related to several true self-avoiding random
walk (\TSAW) models (see \subfig{fig:TSAW}{a-e}, the shorthand
\TSAWfig{a} refers to the model illustrated in
\subfig{fig:TSAW}{a}~\cite{Toth1995}, \etc).
Self-avoiding random walks were originally
introduced~\cite{Amit1983} in order to differentiate self-avoiding lattice
polymers (which are not proper random walks) from \quot{true} random walks that
remember the number of previous passages (called \quot{local  times}) on each
site. Renormalization
arguments~\cite{Pietronero1983} indicate that a one-dimensional \TSAW explores
a region of size $\sim t^{2/3}$ in a time $t$. A variant~\cite{Toth1995} of
the original model records the number of passages across edges (rather than
on sites), allowing for a rigorous analysis. At zero
temperature, on a one-dimensional infinite lattice, this
\TSAWfig{a} has a particle move to either side
with equal probabilities if the two neighboring edges have the same local time,
and otherwise moves across the edge with the smaller of the two local times. 
It was proved for this model that for positive
temperature, during a time $t$, the particle explores a region of size
$t^{2/3}$~\cite{Toth1995}. Unlike the random walk of the pointer itself,
the combined evolution of
the pointer and the local times is a Markov chain.
 Clearly, at zero temperature, this Markov chain requires initial
non-constant local times in order to avoid pathological behavior. In this
setting, the  $t^{2/3}$ scaling likely still holds, so that $t^{2/3} = L$ agrees
with the $3/2$ exponent of \eq{equ:RelaxationTime}. In \REF[Sect.
11]{TothWerner1998}, it is shown that \TSAWfig{a} is equivalent to
\TSAWfig{b},
namely a lattice model of particles and holes, where a single site contains a
pair of items (either  two
particles, two holes, or a particle and a hole). In  \TSAWfig{b}, at each
time step, one of the
two paired items, chosen randomly, moves to the left if it is a hole and to
the right if it is a particle.

The models \TSAWfig{a,b}  are furthermore
equivalent to the \VLT (see \subfig{fig:TSAW}{c})~\cite{Erignoux2025b}, defined
by the
transitions
\begin{align}
\RIGHT \PART & \rightarrow \PART \RIGHT \quad \quad  p=1 \label{equ:Fleches1} \\
\HOLE\LEFT   & \rightarrow \LEFT \HOLE \quad \quad p=1 \label{equ:Fleches2}
\end{align}
and
\begin{equation}
 \begin{rcases}
 \RIGHT \HOLE \\
 \PART \LEFT
 \end{rcases}
 \to
 \begin{cases}
\LEFT \PART& p = \alpha\\
\HOLE \RIGHT & p = 1-\alpha.
 \end{cases}
 \label{equ:Fleches3}
\end{equation}
The configurations of the \VLT have the same trajectories as in the \LT with a 
time change. In the \VLT, when the pointer encounters a cluster of particles, it 
crosses it deterministically by \eq{equ:Fleches1}, whereas in the \LT, by
\eq{equ:LTASEP2}, this cluster is ultimately crossed but in a typically
longer random time. Conversely, in the \LT, after a pullback move 
of \eq{equ:LTASEP1}, the pointer crosses instantly the empty zone behind it,
while in the \VLT, this move is broken down into deterministic local
steps of \eq{equ:Fleches2}.
We can thus justify ~\cite{Erignoux2025b} that the two models have
overall the same equilibrium and out-of-equilibrium time scales.
This establishes the equivalence of the \LT with the \TSAWfig{a}  on an
infinite lattice.
On a lattice of length $L$ with periodic
boundary conditions, the \VLT converges for all  $\alpha$ to the stationary
state with equal weights for all configurations. Finally, we consider the
zero-temperature \quot{oriented} \TSAWfig{d}, in which each edge records
directional local
times~\cite{Veto2008}. In order to jump away from a site, the two outgoing local times are
compared. Started from all-zero local times,  \TSAWfig{d} is also
equivalent  to the \VLT with a step initial
configuration \cite{Erignoux2025b}. For \TSAWfig{d}, the scaling
$\critsize{t}
\simeq 2 \sqrt{t}$ and the uniform position of the pointer inside $\crit{t}$ are
evidenced  in \REF{Veto2008}, which further justifies \eq{equ:Dome10bis}. 
The \VLT thus connects the original \TSAWfig{a,b}
\cite{Toth1995,TothWerner1998}
and its oriented variant  \TSAWfig{d}~\cite{Veto2008}, the former being an
equilibrium version of the latter. The two variants map onto two different regimes of
the \LT, namely an equilibrium-like and a  step setting. 
The $t^{2/3}$ scaling of the pointer motion at zero temperature
is supported by its continuous limit  derived in \REF{Erignoux2025b}, 
which yields the equation of the  
true self-avoiding motion \cite{TothWerner1998}, for which the $2/3$ exponent is 
proven.

In conclusion, the self-trapping of velocities in non-reversible Markov chains,
that we described here for the \LT at the critical pullback $\alphacrit$,
originates in velocity--density correlation, and it generates exceptionally fast
local dynamics, both out of equilibrium and in equilibrium. The phenomenon
extends to generic one-dimensional particle models on the
lattice~\cite{Essler2025} or, under event-chain Monte Carlo dynamics, to the
continuum~\cite{Lei2019,Krauth2024hamiltonian}. It has also been
demonstrated in higher-dimensional models~\cite{MaggsKrauth2022}. The \LT
departs from the logic of gradient-based simulation methods, for example in
the discussed dome-shaped expansion of the critical region. Particles far
outside the dome are strictly arrested---they never receive the pointer. In
consequence, they do not add to the computational burden for equilibrating the
system. Non-equilibrium and equilibrium scalings are thus faster than for
the (unlifted) \TASEP and for Hamiltonian Monte Carlo. Remarkably, the \LUT time
evolution is totally unbiased, and Metropolis corrections and resamplings as
in Hamiltonian Monte Carlo are not needed. We expect the principles
at work in the \LT to be of more general use in optimization and molecular
simulation, where gradient-based and stochastic gradient-based methods are
widespread but likely not optimal.

\begin{acknowledgments}
We thank F. H. L. Essler, K. Hukushima, A. C. Maggs, S. Todo, and B. Tóth for
helpful discussions. Research of W. K. was supported by a grant from the Simons
Foundation (Grant 839534, MET). W. K. thanks the Isaac Newton Institute for
Mathematical Sciences, Cambridge, for support and hospitality during the
programme Monte Carlo sampling: beyond the diffusive regime, where work on this
paper was undertaken. This work was supported by EPSRC grant EP/Z000580/1. 
This work was supported by PSL via the GP Statistical Physics and Mathematics.
\end{acknowledgments}


\begin{thebibliography}{32}%
\makeatletter
\providecommand \@ifxundefined [1]{%
 \@ifx{#1\undefined}
}%
\providecommand \@ifnum [1]{%
 \ifnum #1\expandafter \@firstoftwo
 \else \expandafter \@secondoftwo
 \fi
}%
\providecommand \@ifx [1]{%
 \ifx #1\expandafter \@firstoftwo
 \else \expandafter \@secondoftwo
 \fi
}%
\providecommand \natexlab [1]{#1}%
\providecommand \enquote  [1]{``#1''}%
\providecommand \bibnamefont  [1]{#1}%
\providecommand \bibfnamefont [1]{#1}%
\providecommand \citenamefont [1]{#1}%
\providecommand \href@noop [0]{\@secondoftwo}%
\providecommand \href [0]{\begingroup \@sanitize@url \@href}%
\providecommand \@href[1]{\@@startlink{#1}\@@href}%
\providecommand \@@href[1]{\endgroup#1\@@endlink}%
\providecommand \@sanitize@url [0]{\catcode `\\12\catcode `\$12\catcode
  `\&12\catcode `\#12\catcode `\^12\catcode `\_12\catcode `\%12\relax}%
\providecommand \@@startlink[1]{}%
\providecommand \@@endlink[0]{}%
\providecommand \url  [0]{\begingroup\@sanitize@url \@url }%
\providecommand \@url [1]{\endgroup\@href {#1}{\urlprefix }}%
\providecommand \urlprefix  [0]{URL }%
\providecommand \Eprint [0]{\href }%
\providecommand \doibase [0]{https://doi.org/}%
\providecommand \selectlanguage [0]{\@gobble}%
\providecommand \bibinfo  [0]{\@secondoftwo}%
\providecommand \bibfield  [0]{\@secondoftwo}%
\providecommand \translation [1]{[#1]}%
\providecommand \BibitemOpen [0]{}%
\providecommand \bibitemStop [0]{}%
\providecommand \bibitemNoStop [0]{.\EOS\space}%
\providecommand \EOS [0]{\spacefactor3000\relax}%
\providecommand \BibitemShut  [1]{\csname bibitem#1\endcsname}%
\let\auto@bib@innerbib\@empty
\bibitem [{\citenamefont {Brush}(1976)}]{Brush1976}%
  \BibitemOpen
  \bibfield  {author} {\bibinfo {author} {\bibfnamefont {S.~G.}\ \bibnamefont
  {Brush}},\ }\href@noop {} {\emph {\bibinfo {title} {{The Kind of Motion We
  Call Heat: A History of the Kinetic Theory of Gases in the 19th Century}}}}\
  (\bibinfo  {publisher} {North-Holland Publishing Company},\ \bibinfo {year}
  {1976})\BibitemShut {NoStop}%
\bibitem [{\citenamefont {Schlick}(2002)}]{SchlickBook}%
  \BibitemOpen
  \bibfield  {author} {\bibinfo {author} {\bibfnamefont {T.}~\bibnamefont
  {Schlick}},\ }\href@noop {} {\emph {\bibinfo {title} {{Molecular Modeling and
  Simulation: An Interdisciplinary Guide}}}}\ (\bibinfo  {publisher}
  {Springer-Verlag},\ \bibinfo {year} {2002})\BibitemShut {NoStop}%
\bibitem [{\citenamefont {Frenkel}\ and\ \citenamefont
  {Smit}(2001)}]{FrenkelSmitBook2001}%
  \BibitemOpen
  \bibfield  {author} {\bibinfo {author} {\bibfnamefont {D.}~\bibnamefont
  {Frenkel}}\ and\ \bibinfo {author} {\bibfnamefont {B.}~\bibnamefont {Smit}},\
  }\href {https://books.google.de/books?id=5qTzldS9ROIC} {\emph {\bibinfo
  {title} {{Understanding Molecular Simulation: From Algorithms to
  Applications}}}},\ Computational science series\ (\bibinfo  {publisher}
  {Elsevier Science},\ \bibinfo {year} {2001})\BibitemShut {NoStop}%
\bibitem [{\citenamefont {Ciccotti}\ \emph {et~al.}(2022)\citenamefont
  {Ciccotti}, \citenamefont {Dellago}, \citenamefont {Ferrario}, \citenamefont
  {Hern{\'a}ndez},\ and\ \citenamefont {Tuckerman}}]{Ciccotti2022}%
  \BibitemOpen
  \bibfield  {author} {\bibinfo {author} {\bibfnamefont {G.}~\bibnamefont
  {Ciccotti}}, \bibinfo {author} {\bibfnamefont {C.}~\bibnamefont {Dellago}},
  \bibinfo {author} {\bibfnamefont {M.}~\bibnamefont {Ferrario}}, \bibinfo
  {author} {\bibfnamefont {E.~R.}\ \bibnamefont {Hern{\'a}ndez}},\ and\
  \bibinfo {author} {\bibfnamefont {M.~E.}\ \bibnamefont {Tuckerman}},\ }\href
  {https://doi.org/10.1140/epjb/s10051-021-00249-x} {\bibfield  {journal}
  {\bibinfo  {journal} {Eur. Phys. J. B}\ }\textbf {\bibinfo {volume} {95}},\
  \bibinfo {pages} {3} (\bibinfo {year} {2022})}\BibitemShut {NoStop}%
\bibitem [{\citenamefont {Duane}\ \emph {et~al.}(1987)\citenamefont {Duane},
  \citenamefont {Kennedy}, \citenamefont {Pendleton},\ and\ \citenamefont
  {Roweth}}]{Duane1987}%
  \BibitemOpen
  \bibfield  {author} {\bibinfo {author} {\bibfnamefont {S.}~\bibnamefont
  {Duane}}, \bibinfo {author} {\bibfnamefont {A.}~\bibnamefont {Kennedy}},
  \bibinfo {author} {\bibfnamefont {B.~J.}\ \bibnamefont {Pendleton}},\ and\
  \bibinfo {author} {\bibfnamefont {D.}~\bibnamefont {Roweth}},\ }\href
  {https://doi.org/https://doi.org/10.1016/0370-2693(87)91197-X} {\bibfield
  {journal} {\bibinfo  {journal} {Phys. Lett. B}\ }\textbf {\bibinfo {volume}
  {195}},\ \bibinfo {pages} {216} (\bibinfo {year} {1987})}\BibitemShut
  {NoStop}%
\bibitem [{\citenamefont {Neal}(2011)}]{Neal2011}%
  \BibitemOpen
  \bibfield  {author} {\bibinfo {author} {\bibfnamefont {R.~M.}\ \bibnamefont
  {Neal}},\ }in\ \href@noop {} {\emph {\bibinfo {booktitle} {{Handbook of
  Markov Chain Monte Carlo}}}},\ \bibinfo {editor} {edited by\ \bibinfo
  {editor} {\bibfnamefont {S.}~\bibnamefont {Brooks}}, \bibinfo {editor}
  {\bibfnamefont {A.}~\bibnamefont {Gelman}}, \bibinfo {editor} {\bibfnamefont
  {G.}~\bibnamefont {Jones}},\ and\ \bibinfo {editor} {\bibfnamefont {X.-L.}\
  \bibnamefont {Meng}}}\ (\bibinfo  {publisher} {Chapman and Hall/CRC},\
  \bibinfo {year} {2011})\ pp.\ \bibinfo {pages} {113--162}\BibitemShut
  {NoStop}%
\bibitem [{\citenamefont {Krauth}(2006)}]{SMAC}%
  \BibitemOpen
  \bibfield  {author} {\bibinfo {author} {\bibfnamefont {W.}~\bibnamefont
  {Krauth}},\ }\href@noop {} {\emph {\bibinfo {title} {{Statistical Mechanics:
  Algorithms and Computations}}}}\ (\bibinfo  {publisher} {Oxford University
  Press},\ \bibinfo {year} {2006})\BibitemShut {NoStop}%
\bibitem [{\citenamefont {Bernard}\ \emph {et~al.}(2009)\citenamefont
  {Bernard}, \citenamefont {Krauth},\ and\ \citenamefont
  {Wilson}}]{Bernard2009}%
  \BibitemOpen
  \bibfield  {author} {\bibinfo {author} {\bibfnamefont {E.~P.}\ \bibnamefont
  {Bernard}}, \bibinfo {author} {\bibfnamefont {W.}~\bibnamefont {Krauth}},\
  and\ \bibinfo {author} {\bibfnamefont {D.~B.}\ \bibnamefont {Wilson}},\
  }\href {https://doi.org/10.1103/PhysRevE.80.056704} {\bibfield  {journal}
  {\bibinfo  {journal} {Phys. Rev. E}\ }\textbf {\bibinfo {volume} {80}},\
  \bibinfo {pages} {056704} (\bibinfo {year} {2009})}\BibitemShut {NoStop}%
\bibitem [{\citenamefont {{Michel}}\ \emph {et~al.}(2014)\citenamefont
  {{Michel}}, \citenamefont {{Kapfer}},\ and\ \citenamefont
  {{Krauth}}}]{Michel2014JCP}%
  \BibitemOpen
  \bibfield  {author} {\bibinfo {author} {\bibfnamefont {M.}~\bibnamefont
  {{Michel}}}, \bibinfo {author} {\bibfnamefont {S.~C.}\ \bibnamefont
  {{Kapfer}}},\ and\ \bibinfo {author} {\bibfnamefont {W.}~\bibnamefont
  {{Krauth}}},\ }\href {https://doi.org/10.1063/1.4863991} {\bibfield
  {journal} {\bibinfo  {journal} {J. Chem. Phys.}\ }\textbf {\bibinfo {volume}
  {140}},\ \bibinfo {eid} {054116} (\bibinfo {year} {2014})}\BibitemShut
  {NoStop}%
\bibitem [{\citenamefont {Krauth}(2021)}]{Krauth2021eventchain}%
  \BibitemOpen
  \bibfield  {author} {\bibinfo {author} {\bibfnamefont {W.}~\bibnamefont
  {Krauth}},\ }\href {https://doi.org/10.3389/fphy.2021.663457} {\bibfield
  {journal} {\bibinfo  {journal} {Front. Phys.}\ }\textbf {\bibinfo {volume}
  {9}},\ \bibinfo {pages} {229} (\bibinfo {year} {2021})}\BibitemShut {NoStop}%
\bibitem [{\citenamefont {Diaconis}\ \emph {et~al.}(2000)\citenamefont
  {Diaconis}, \citenamefont {Holmes},\ and\ \citenamefont
  {Neal}}]{Diaconis2000}%
  \BibitemOpen
  \bibfield  {author} {\bibinfo {author} {\bibfnamefont {P.}~\bibnamefont
  {Diaconis}}, \bibinfo {author} {\bibfnamefont {S.}~\bibnamefont {Holmes}},\
  and\ \bibinfo {author} {\bibfnamefont {R.~M.}\ \bibnamefont {Neal}},\ }\href
  {https://doi.org/10.1214/aoap/1019487508} {\bibfield  {journal} {\bibinfo
  {journal} {Ann. Appl. Probab.}\ }\textbf {\bibinfo {volume} {10}},\ \bibinfo
  {pages} {726} (\bibinfo {year} {2000})}\BibitemShut {NoStop}%
\bibitem [{\citenamefont {Chen}\ \emph {et~al.}(1999)\citenamefont {Chen},
  \citenamefont {Lovász},\ and\ \citenamefont {Pak}}]{Chen1999}%
  \BibitemOpen
  \bibfield  {author} {\bibinfo {author} {\bibfnamefont {F.}~\bibnamefont
  {Chen}}, \bibinfo {author} {\bibfnamefont {L.}~\bibnamefont {Lovász}},\ and\
  \bibinfo {author} {\bibfnamefont {I.}~\bibnamefont {Pak}},\ }\href@noop {}
  {\bibfield  {journal} {\bibinfo  {journal} {Proceedings of the 17th Annual
  ACM Symposium on Theory of Computing}\ ,\ \bibinfo {pages} {275}} (\bibinfo
  {year} {1999})}\BibitemShut {NoStop}%
\bibitem [{\citenamefont {Bernard}\ and\ \citenamefont
  {Krauth}(2011)}]{Bernard2011}%
  \BibitemOpen
  \bibfield  {author} {\bibinfo {author} {\bibfnamefont {E.~P.}\ \bibnamefont
  {Bernard}}\ and\ \bibinfo {author} {\bibfnamefont {W.}~\bibnamefont
  {Krauth}},\ }\href {https://doi.org/10.1103/PhysRevLett.107.155704}
  {\bibfield  {journal} {\bibinfo  {journal} {Phys. Rev. Lett.}\ }\textbf
  {\bibinfo {volume} {107}},\ \bibinfo {pages} {155704} (\bibinfo {year}
  {2011})}\BibitemShut {NoStop}%
\bibitem [{\citenamefont {Kampmann}\ \emph {et~al.}(2021)\citenamefont
  {Kampmann}, \citenamefont {M\"{u}ller}, \citenamefont {Weise}, \citenamefont
  {Vorsmann},\ and\ \citenamefont {Kierfeld}}]{Kampmann2021}%
  \BibitemOpen
  \bibfield  {author} {\bibinfo {author} {\bibfnamefont {T.~A.}\ \bibnamefont
  {Kampmann}}, \bibinfo {author} {\bibfnamefont {D.}~\bibnamefont
  {M\"{u}ller}}, \bibinfo {author} {\bibfnamefont {L.~P.}\ \bibnamefont
  {Weise}}, \bibinfo {author} {\bibfnamefont {C.~F.}\ \bibnamefont
  {Vorsmann}},\ and\ \bibinfo {author} {\bibfnamefont {J.}~\bibnamefont
  {Kierfeld}},\ }\href {https://doi.org/10.3389/fphy.2021.635886} {\bibfield
  {journal} {\bibinfo  {journal} {Front. Phys.}\ }\textbf {\bibinfo {volume}
  {9}},\ \bibinfo {pages} {96} (\bibinfo {year} {2021})}\BibitemShut {NoStop}%
\bibitem [{\citenamefont {Klement}\ and\ \citenamefont
  {Engel}(2019)}]{Klement2019}%
  \BibitemOpen
  \bibfield  {author} {\bibinfo {author} {\bibfnamefont {M.}~\bibnamefont
  {Klement}}\ and\ \bibinfo {author} {\bibfnamefont {M.}~\bibnamefont
  {Engel}},\ }\href {https://doi.org/10.1063/1.5090882} {\bibfield  {journal}
  {\bibinfo  {journal} {J. Chem. Phys.}\ }\textbf {\bibinfo {volume} {150}},\
  \bibinfo {pages} {174108} (\bibinfo {year} {2019})}\BibitemShut {NoStop}%
\bibitem [{\citenamefont {Maggs}(2024{\natexlab{a}})}]{Maggs2024}%
  \BibitemOpen
  \bibfield  {author} {\bibinfo {author} {\bibfnamefont {A.~C.}\ \bibnamefont
  {Maggs}},\ }\href {https://doi.org/10.1209/0295-5075/ad64ff} {\bibfield
  {journal} {\bibinfo  {journal} {EPL}\ }\textbf {\bibinfo {volume} {147}},\
  \bibinfo {pages} {21001} (\bibinfo {year} {2024}{\natexlab{a}})}\BibitemShut
  {NoStop}%
\bibitem [{\citenamefont {Maggs}(2024{\natexlab{b}})}]{Maggs2024b}%
  \BibitemOpen
  \bibfield  {author} {\bibinfo {author} {\bibfnamefont {A.~C.}\ \bibnamefont
  {Maggs}},\ }\href {https://arxiv.org/abs/2410.08694} {\bibinfo {title}
  {{Event-chain Monte Carlo and the true self-avoiding walk}}} (\bibinfo {year}
  {2024}{\natexlab{b}}),\ \bibinfo {note} {arXiv:2410.08694}\BibitemShut
  {NoStop}%
\bibitem [{\citenamefont {Kapfer}\ and\ \citenamefont
  {Krauth}(2017)}]{KapferKrauth2017}%
  \BibitemOpen
  \bibfield  {author} {\bibinfo {author} {\bibfnamefont {S.~C.}\ \bibnamefont
  {Kapfer}}\ and\ \bibinfo {author} {\bibfnamefont {W.}~\bibnamefont
  {Krauth}},\ }\href {https://doi.org/10.1103/PhysRevLett.119.240603}
  {\bibfield  {journal} {\bibinfo  {journal} {Phys. Rev. Lett.}\ }\textbf
  {\bibinfo {volume} {119}},\ \bibinfo {pages} {240603} (\bibinfo {year}
  {2017})}\BibitemShut {NoStop}%
\bibitem [{\citenamefont {Lei}\ \emph {et~al.}(2019)\citenamefont {Lei},
  \citenamefont {Krauth},\ and\ \citenamefont {Maggs}}]{Lei2019}%
  \BibitemOpen
  \bibfield  {author} {\bibinfo {author} {\bibfnamefont {Z.}~\bibnamefont
  {Lei}}, \bibinfo {author} {\bibfnamefont {W.}~\bibnamefont {Krauth}},\ and\
  \bibinfo {author} {\bibfnamefont {A.~C.}\ \bibnamefont {Maggs}},\ }\bibfield
  {journal} {\bibinfo  {journal} {Phys. Rev. E}\ }\textbf {\bibinfo {volume}
  {99}},\ \href {https://doi.org/10.1103/physreve.99.043301}
  {10.1103/physreve.99.043301} (\bibinfo {year} {2019})\BibitemShut {NoStop}%
\bibitem [{\citenamefont {Amit}\ \emph {et~al.}(1983)\citenamefont {Amit},
  \citenamefont {Parisi},\ and\ \citenamefont {Peliti}}]{Amit1983}%
  \BibitemOpen
  \bibfield  {author} {\bibinfo {author} {\bibfnamefont {D.~J.}\ \bibnamefont
  {Amit}}, \bibinfo {author} {\bibfnamefont {G.}~\bibnamefont {Parisi}},\ and\
  \bibinfo {author} {\bibfnamefont {L.}~\bibnamefont {Peliti}},\ }\href
  {https://doi.org/10.1103/PhysRevB.27.1635} {\bibfield  {journal} {\bibinfo
  {journal} {Phys. Rev. B}\ }\textbf {\bibinfo {volume} {27}},\ \bibinfo
  {pages} {1635} (\bibinfo {year} {1983})}\BibitemShut {NoStop}%
\bibitem [{\citenamefont {Pietronero}(1983)}]{Pietronero1983}%
  \BibitemOpen
  \bibfield  {author} {\bibinfo {author} {\bibfnamefont {L.}~\bibnamefont
  {Pietronero}},\ }\href {https://doi.org/10.1103/PhysRevB.27.5887} {\bibfield
  {journal} {\bibinfo  {journal} {Phys. Rev. B}\ }\textbf {\bibinfo {volume}
  {27}},\ \bibinfo {pages} {5887} (\bibinfo {year} {1983})}\BibitemShut
  {NoStop}%
\bibitem [{\citenamefont {Bernasconi}\ and\ \citenamefont
  {Pietronero}(1984)}]{BernasconiPietronero1984}%
  \BibitemOpen
  \bibfield  {author} {\bibinfo {author} {\bibfnamefont {J.}~\bibnamefont
  {Bernasconi}}\ and\ \bibinfo {author} {\bibfnamefont {L.}~\bibnamefont
  {Pietronero}},\ }\href {https://doi.org/10.1103/PhysRevB.29.5196} {\bibfield
  {journal} {\bibinfo  {journal} {Phys. Rev. B}\ }\textbf {\bibinfo {volume}
  {29}},\ \bibinfo {pages} {5196} (\bibinfo {year} {1984})}\BibitemShut
  {NoStop}%
\bibitem [{\citenamefont {Brémont}\ \emph {et~al.}(2024)\citenamefont
  {Brémont}, \citenamefont {Bénichou},\ and\ \citenamefont
  {Voituriez}}]{Bremont2024}%
  \BibitemOpen
  \bibfield  {author} {\bibinfo {author} {\bibfnamefont {J.}~\bibnamefont
  {Brémont}}, \bibinfo {author} {\bibfnamefont {O.}~\bibnamefont
  {Bénichou}},\ and\ \bibinfo {author} {\bibfnamefont {R.}~\bibnamefont
  {Voituriez}},\ }\bibfield  {journal} {\bibinfo  {journal} {Phys. Rev. Lett.}\
  }\textbf {\bibinfo {volume} {133}},\ \href
  {https://doi.org/10.1103/physrevlett.133.157101}
  {10.1103/physrevlett.133.157101} (\bibinfo {year} {2024})\BibitemShut
  {NoStop}%
\bibitem [{\citenamefont {Toth}(1995)}]{Toth1995}%
  \BibitemOpen
  \bibfield  {author} {\bibinfo {author} {\bibfnamefont {B.}~\bibnamefont
  {Toth}},\ }\href {https://doi.org/10.1214/aop/1176987793} {\bibfield
  {journal} {\bibinfo  {journal} {The Annals of Probability}\ }\textbf
  {\bibinfo {volume} {23}},\ \bibinfo {pages} {1523} (\bibinfo {year}
  {1995})},\ \bibinfo {note} {publisher: Institute of Mathematical
  Statistics}\BibitemShut {NoStop}%
\bibitem [{\citenamefont {Tóth}\ and\ \citenamefont
  {Werner}(1998)}]{TothWerner1998}%
  \BibitemOpen
  \bibfield  {author} {\bibinfo {author} {\bibfnamefont {B.}~\bibnamefont
  {Tóth}}\ and\ \bibinfo {author} {\bibfnamefont {W.}~\bibnamefont {Werner}},\
  }\href {https://doi.org/10.1007/s004400050172} {\bibfield  {journal}
  {\bibinfo  {journal} {Probab. Theory and Relat. Fields}\ }\textbf {\bibinfo
  {volume} {111}},\ \bibinfo {pages} {375–452} (\bibinfo {year}
  {1998})}\BibitemShut {NoStop}%
\bibitem [{\citenamefont {Toth}\ and\ \citenamefont {Veto}(2008)}]{Veto2008}%
  \BibitemOpen
  \bibfield  {author} {\bibinfo {author} {\bibfnamefont {B.}~\bibnamefont
  {Toth}}\ and\ \bibinfo {author} {\bibfnamefont {B.}~\bibnamefont {Veto}},\
  }\bibfield  {journal} {\bibinfo  {journal} {Electron. J. Probab.}\ }\textbf
  {\bibinfo {volume} {13}},\ \href {https://doi.org/10.1214/ejp.v13-570}
  {10.1214/ejp.v13-570} (\bibinfo {year} {2008})\BibitemShut {NoStop}%
\bibitem [{\citenamefont {Dumaz}\ and\ \citenamefont
  {Tóth}(2013)}]{Dumaz2013}%
  \BibitemOpen
  \bibfield  {author} {\bibinfo {author} {\bibfnamefont {L.}~\bibnamefont
  {Dumaz}}\ and\ \bibinfo {author} {\bibfnamefont {B.}~\bibnamefont {Tóth}},\
  }\href {https://doi.org/10.1016/j.spa.2012.11.011} {\bibfield  {journal}
  {\bibinfo  {journal} {Stoch. Process. Their Appl.}\ }\textbf {\bibinfo
  {volume} {123}},\ \bibinfo {pages} {1454–1471} (\bibinfo {year}
  {2013})}\BibitemShut {NoStop}%
\bibitem [{\citenamefont {Essler}\ and\ \citenamefont
  {Krauth}(2024)}]{Essler2024PRX}%
  \BibitemOpen
  \bibfield  {author} {\bibinfo {author} {\bibfnamefont {F.~H.~L.}\
  \bibnamefont {Essler}}\ and\ \bibinfo {author} {\bibfnamefont
  {W.}~\bibnamefont {Krauth}},\ }\href
  {https://doi.org/10.1103/PhysRevX.14.041035} {\bibfield  {journal} {\bibinfo
  {journal} {Phys. Rev. X}\ }\textbf {\bibinfo {volume} {14}},\ \bibinfo
  {pages} {041035} (\bibinfo {year} {2024})}\BibitemShut {NoStop}%
\bibitem [{\citenamefont {Erignoux}\ \emph {et~al.}(2025)\citenamefont
  {Erignoux}, \citenamefont {Krauth}, \citenamefont {Massoulié},\ and\
  \citenamefont {Toninelli}}]{Erignoux2025b}%
  \BibitemOpen
  \bibfield  {author} {\bibinfo {author} {\bibfnamefont {C.}~\bibnamefont
  {Erignoux}}, \bibinfo {author} {\bibfnamefont {W.}~\bibnamefont {Krauth}},
  \bibinfo {author} {\bibfnamefont {B.}~\bibnamefont {Massoulié}},\ and\
  \bibinfo {author} {\bibfnamefont {C.}~\bibnamefont {Toninelli}}} (\bibinfo
  {year} {2025}),\ \bibinfo {note} {manuscript in preparation}\BibitemShut
  {NoStop}%
\bibitem [{\citenamefont {Krauth}(2024)}]{Krauth2024hamiltonian}%
  \BibitemOpen
  \bibfield  {author} {\bibinfo {author} {\bibfnamefont {W.}~\bibnamefont
  {Krauth}},\ }\href {https://arxiv.org/abs/2411.11690} {\bibinfo {title}
  {{Hamiltonian Monte Carlo vs. event-chain Monte Carlo: an appraisal of
  sampling strategies beyond the diffusive regime}}} (\bibinfo {year} {2024}),\
  \Eprint {https://arxiv.org/abs/2411.11690} {arXiv:2411.11690
  [cond-mat.stat-mech]} \BibitemShut {NoStop}%
\bibitem [{\citenamefont {Essler}\ \emph {et~al.}(2025)\citenamefont {Essler},
  \citenamefont {Gipouloux},\ and\ \citenamefont {Krauth}}]{Essler2025}%
  \BibitemOpen
  \bibfield  {author} {\bibinfo {author} {\bibfnamefont {F.~H.~L.}\
  \bibnamefont {Essler}}, \bibinfo {author} {\bibfnamefont {J.}~\bibnamefont
  {Gipouloux}},\ and\ \bibinfo {author} {\bibfnamefont {W.}~\bibnamefont
  {Krauth}},\ }\href {https://arxiv.org/abs/2502.16549} {\bibinfo {title}
  {{Lifted TASEP: long-time dynamics, generalizations, and continuum limit}}}
  (\bibinfo {year} {2025}),\ \Eprint {https://arxiv.org/abs/2502.16549}
  {arXiv:2502.16549 [cond-mat.stat-mech]} \BibitemShut {NoStop}%
\bibitem [{\citenamefont {Maggs}\ and\ \citenamefont
  {Krauth}(2022)}]{MaggsKrauth2022}%
  \BibitemOpen
  \bibfield  {author} {\bibinfo {author} {\bibfnamefont {A.~C.}\ \bibnamefont
  {Maggs}}\ and\ \bibinfo {author} {\bibfnamefont {W.}~\bibnamefont {Krauth}},\
  }\href {https://doi.org/10.1103/PhysRevE.105.015309} {\bibfield  {journal}
  {\bibinfo  {journal} {Phys. Rev. E}\ }\textbf {\bibinfo {volume} {105}},\
  \bibinfo {pages} {015309} (\bibinfo {year} {2022})}\BibitemShut {NoStop}%
\end{thebibliography}
\end{document}